\documentclass[%
 reprint,
 amsmath,amssymb,
 aps,
]{revtex4-2}

\usepackage{ulem}
\usepackage{graphicx}
\usepackage{comment}
\usepackage{dcolumn}
\usepackage{bm}

\usepackage[compat=1.1.0]{tikz-feynman}

\usepackage{bbm}
\newcommand{\Nt}{\ensuremath{\widetilde{N}}}
\newcommand{\Mt}{\ensuremath{\widetilde{M}}}

\begin{document}

\preprint{APS/123-QED}

\title{Percolation properties of the neutron population in nuclear reactors}

\author{Benjamin Dechenaux}
 \email{benjamin.dechenaux@irsn.fr}
\author{Thomas Delcambre}%

\affiliation{%
Institut de Radioprotection et de Sûreté Nucl\'eaire (IRSN)\\
 PSN-RES/SNC/LN, F-92260, Fontenay-aux-Roses, France. 
}%

\author{Eric Dumonteil}
\affiliation{
 Institut de Recherche sur les Lois Fondamentales de l'Univers\\
 CEA, Université Paris-Saclay, 91191 Gif-sur-Yvette, France
}%

\date{\today}

\begin{abstract}
 
\begin{description}
\item[Abstract]
Reactor physics aims at studying the neutron population in a reactor core under the influence of feedback mechanisms, such as the Doppler temperature effect. Numerical schemes to calculate macroscopic properties emerging from such coupled stochastic systems however require to define intermediate quantities (e.g. the temperature field), which are bridging the gap between the stochastic neutron field and the deterministic feedback. 
By interpreting the branching random walk of neutrons in fissile media under the influence of a feedback mechanism as a directed percolation process and by leveraging on the statistical field theory of birth death processes, we will build a stochastic model of neutron transport theory and of reactor physics. The critical exponents of this model, combined to the analysis of the resulting field equation involving a fractional Laplacian will show that the critical diffusion equation cannot adequately describe the spatial distribution of the neutron population and shifts instead to a critical super-diffusion equation. The analysis of this equation will reveal that non-negligible departure from mean field behavior might develop in reactor cores, questioning the attainable accuracy of the numerical schemes currently used by the nuclear industry. 
\end{description}
\end{abstract}

\maketitle


\section{\label{sec:intro} Introduction }

At their inception in the 30's, Monte Carlo algorithms intended to describe the neutron transport in fissile media \cite{Spanier1969MonteProblems1,Lux1991MonteCalculations} in support of the first applications of nuclear power, might it be for the design of the first nuclear weapons or in support of nuclear energy production. Since then, the nuclear energy industry has used the so-called Monte Carlo criticality codes as reference numerical schemes to solve the linear Boltzmann equation. Nuclear safety demonstrations in particular rely on a Monte Carlo solving of this equation with few, if any, hypotheses, while the use of deterministic codes is preferred for an approximate but fast -and even online- solving of a simplified version of this equation: the two (energy) group critical diffusion equation. In both cases, the aim of such neutronics codes \footnote{neutron transport theory, also known as neutronics, aims at characterizing the properties of the neutron gas in fissile media} is to calculate the neutron spatial distribution in the context of mean-field hypotheses. Indeed, even at the startup of a nuclear reactor (i.e. when the neutron densities are small), such mean-field equations are believed to accurately describe the average behavior of the stochastic regime. Though, in the power regime of the reactor, stabilizing effects come into action. For example, local power excursions and potential sudden rise of local neutron densities are tempered by the Doppler broadening of the neutron cross-sections (giving the probabilities that a neutron at a given energy induce a particular reaction on a nuclei): the quantum resonances of the neutron-nuclei system are broadened by the thermal agitation of the nuclei from which results a modification of the neutron capture cross section. The study of the properties of neutron transport in fissile media subject to feedback mechanisms is called 'reactor physics'. Reactor physics numerical solvers hence couple the mean-field equations of neutronics to thermal or thermal-hydraulics solvers via the definition of intermediate quantities (such as the temperature scalar field) to extract macroscopic measurable quantities characterizing the neutron population (such as local neutron fluxes measured by fission rate chambers for example). 

The fluctuations of the neutron population were also largely investigated in the past fifty years \cite{bell_glasstone,williams_book,paszit_pal}, while being assumed to be of marginal concern for nuclear safety, since they vanish as the core power increases. However, a few years ago, this paradigm was questioned \cite{Dumonteil2014ParticleSimulations,Zoia2014ClusteringGeometries,Sutton2017NeutronCalculations} by the study of a neutronics toy-model based on the celebrated branching Brownian motion \cite{harris_book}, that couples the diffusive random walk of neutrons \cite{Zoia2012BranchingStatistics}, mimicking their transport in the Brownian regime, to a Galton-Watson birth-death process, reproducing the induced fission reaction with variable number of outgoing neutrons. Indeed, the first moments of the master equation of this model revealed that the spatial correlations within the neutron population might in some cases diverge, invalidating the use of the mean-field equation. From a physical point of view, this clustering phenomenon (already characterized at the time in the theory of population ecology \cite{Young2001ReproductiveOrganisms, Houchmandzadeh2002ClusteringOrganisms, Houchmandzadeh2008NeutralCommunity}) causes the emergence of spatial patterns in the neutron gas. While always present and increasing with time in 1 and 2 dimensional systems, the spatial correlations however saturate in 3 dimensions. A dedicated experiment taking place at the Rensselaer Poytechnic Institute was designed and took place in 2018 to characterize such spatio-temporal correlations affecting the neutron distributions. This experiment revealed the existence of such correlations and triggered questions relative to their persistence in the power regime of reactor physics where feedbacks occur \cite{Dumonteil2021PatchyReactions}.\newline

In this paper, leveraging on the seminal approach of Janssen and De Dominicis \cite{janssen1976,dominicis1976}, we will build a simplified model of reactor physics to investigate the validity of mean field equations in the power regime of nuclear reactors. In that aim, a field theoretic formulation of neutron physics will be built in Section 2. While being 0 dimensional, this formulation will then be exploited in Section 3 so as to take into account the principal features of neutronics such as the calculation of fluctuations and correlations of the neutron field or the effects of an external neutron source. It will also be extended in order to add the effect of delayed neutrons in Section 4 and to take into account spatial phenomena in Section 5. Along the path, we will highlight the close relationship between  the response functional formalism of Janssen and De Dominics with the Doi and Peliti field theory \cite{doi_1976, peliti_1985} which lend itself to an exact microscopic interpretation, thus strengthening the use of both approaches in neutronics. Finally, in Section 6, we will add a simplified model of thermal feedback, showing that the neutron gas in a nuclear reactor operated at criticality in the power regime can be described as a phase transition belonging to the directed percolation universality class. The calculation of the critical exponents at the percolation threshold using the renormalization group will in return allow to specify the mean field equation, formulated in terms of fractional Laplacian. We will in particular show that in the directed percolation model, the critical diffusion equation shifts to a critical super-diffusion equation, whose stochastic generators are Lévy flights, ultimately dynamically modifying the medium properties in which neutrons are propagating. The analysis of the fundamental mode of the associated fractional Laplacian using an approximate formula will allow to quantify the differences between this formal approach and classical coupling schemes, thus setting an accuracy limit on all actual numerical solvers of reactor physics.

\section{ A Field theoretic formulation of neutronics}
\label{sec:FieldFormulation}
Branching processes are among the simplest models capable of characterizing the phenomenology of neutrons evolving in fissile materials \cite{paszit_pal,williams_book}. In their simplest expression, these models approximate the path of neutrons in matter as a stochastic process, in which they are subject to random extinction or reproduction events, occurring at constant average rates. They are also known to display a second order phase transition between an active and an inactive state \cite{harris_book}. This transition is at the core root of nuclear reactors operations and corresponds, in this context, to the appearance of self-sustained chain reactions: the 'percolation threshold' where neutron chains survive indefinitely is precisely the point at which nuclear reactors do operate. 

Field theoretic method have long been used in statistical and condensed matter physics for their efficiency in the study of the intricate behavior of systems undergoing phase transitions. They indeed prove particularly adapted to unravel the universal properties and scaling behaviors in the vicinity of the phase transition's critical point. Along this line of thought, the present paper ought to demonstrate that field theory also provides an adapted framework for neutronics, allowing notably to unravel highly non-trivial behaviors of the neutron population close to \textit{criticality}.

The derivation of a field-theoretic version of neutronics followed in this paper is based on the seminal work of Janssen \cite{janssen1976} and De Dominicis \cite{dominicis1976}. Starting from the mean field equation of the problem, to which is supplemented a random noise source term (as is commonly done to study the stability of nuclear reactors \cite{bell_glasstone}), the neutron population evolving in a nuclear reactor can be decoupled into two components:
\begin{itemize}
    \item a smoothly varying neutron field $N(t)$. It corresponds to the instantaneous number of neutrons that can be measured in a reactor (proportional to the operating power for instance) and it is only sensitive to the global reactivity $\rho$ of the reactor;
    
    \item a random noise source $\eta(t)$, arising from various perturbations (such as the stochasticity of the induced fission reactions that have a random number of outgoing neutrons, or such as assembly vibrations) which are generally driven by unknown or at least unspecified phenomena. 
\end{itemize}

In the simplest model of the branching process associated to neutrons evolving in an idealized, infinite and homogeneous reactor, such a decoupling transforms the mean field equation into a Stochastic Differential Equation (SDE), whose form is given by  
\begin{equation}
\label{LangevinEquation}
    \cfrac{d N(t)}{dt} = \rho N(t) + \eta(t).
\end{equation}

Any macroscopic observable quantity $\mathcal{O}[N]$ is insensitive to the rapidly varying and low lying noise term. As such, physical observables must be averaged over all possible configurations of the noise. This is formally achieved by performing a functional integration over the noise probability functional $P[\eta(t)]$,
\begin{equation}
\label{observable}
    \langle \mathcal{O}[N] \rangle \, \propto \, \int \mathcal{D}\eta \, \mathcal{O}[N] \, P[\eta].
\end{equation}

The development of the field-theoretic model starts by noting that the $N$ appearing under the integral sign of Eq.~\ref{observable} is constrained to be a solution of Eq.~\ref{LangevinEquation}. The constraint (valid for all $t$),
\begin{equation}
    \label{constraint_N}
    \mathcal{C}[N] \, = \, \left\{\frac{dN(t)}{dt} \, - \, \rho \, N(t) \,  -  \, \eta(t)\right\} = 0,
\end{equation}
can be enforced with the help of a functional version of the resolution of the identity \cite{cardy_tauber},
\begin{equation}
    1 = \int \mathcal{D}N \, \prod_t  \delta \left( \mathcal{C}[N] \right).
\end{equation}
Performing a (functional) Fourier transform, one can write
\begin{eqnarray}
\label{constrained_field}
    1 & = & \int \mathcal{D}[i\Nt] \int \mathcal{D}N \, e^{- \int dt \Nt(t) \, \mathcal{C}[N]} \nonumber\\
    & = & \int \mathcal{D}[i\Nt] \int \mathcal{D}N \, e^{- \int dt \Nt(t) \, \left( \frac{dN(t)}{dt} \, - \, \rho \, N(t) \,  -  \, \eta(t)\right)},
\end{eqnarray}
where the purely imaginary auxiliary field $\Nt(t)$, known as the Martin-Siggia-Rose response field \cite{martin_siggia_rose}, has been introduced. Inserting this Fourier transformed identity back into Eq.~\ref{observable}, one obtains
\begin{eqnarray}
\label{observable2}
    \langle \mathcal{O}[N] \rangle & \propto & \int \mathcal{D}[i\Nt] \int \mathcal{D}N \left\{ e^{- \int \Nt(t)  \left( \frac{d}{dt}  -  \rho   \right)N(t) \, dt} \right. \nonumber \\ 
    & \times &\mathcal{O}[N] 
   \left.  \times  \int \mathcal{D}\eta \, e^{\int \Nt(t) \eta(t) dt} P[\eta] \right\}.
\end{eqnarray}
To go further, one needs to integrate over $\eta$ and thus specify the probability functional to which it is associated. In the following, it will be supposed that at any given moment in time, the noise follows a Gaussian probability distribution 
\begin{equation}
\label{noise_proba}
    P[\eta] \propto e^{ -\frac{1}{4}\int \eta(t) \Gamma^{-1} \eta(t) \, dt }
\end{equation}
with zero mean $\langle \eta(t) \rangle = 0$ and a covariance given by, 
\begin{equation}
    \langle \eta(t) \eta(t') \rangle = 2 \Gamma(N) \delta(t - t').
\end{equation} 
Note that an explicit dependence of the variance $\Gamma(N)$ on the global power level (i.e. $N(t)$) is to be expected on physical ground : the expected noise in a reactor indeed depends on the operating power level. The integral over the noise can now be straightforwardly evaluated as
\begin{equation}
\int \mathcal{D}\eta \, e^{\int \Nt(t) \eta(t) dt} P[\eta] \propto e^{ \Gamma \Nt^2}.
\end{equation}
Inserting this result back into Eq.~\ref{observable2}, one finally ar\-rives at an expression that solely depends on the variance of the noise probability distribution
\begin{equation}
\label{observable_final}
    \langle \mathcal{O}[N] \rangle \, = \mathcal{N}  \,  \int \mathcal{D}[i\Nt] \int \mathcal{D}N \, \mathcal{O}[N] \, 
    e^{- \mathcal{S}[N,\Nt]},
\end{equation}
where both a normalisation factor $\mathcal{N}$ and the so-called \textit{response functional} have been introduced, this last functional being defined through
\begin{equation}
\label{response_functional}
     \mathcal{S}[N,\Nt] = \int \left\{\Nt(t)  \left( \frac{d}{dt} - \rho \right)N(t) - \Nt(t) \Gamma \Nt(t)\, \right\} dt.
\end{equation}

It often reveals convenient to include in Eq.~\ref{constraint_N} an initial condition of the type $N(t_0) = N_0$. The initial condition is then included directly into the response functional and it will be taken into account naturally in the further calculation of observables. One can show that the inclusion of an initial condition in the response functional gives
\begin{eqnarray}
\label{response_functional_initial_cond}
     \mathcal{S}[N,\Nt]  & \, & = \int \left\{  \Nt(t)  \left( \frac{d}{dt} - \rho \right)N(t)          \right.\nonumber\\
      && \left. -  \Nt(t) \, \delta(t-t_0) \, N_0   -   \Nt(t) \Gamma \Nt(t)\, \right\} dt.
\end{eqnarray}
It can then be demonstrated that the net effect of this supplementary term appearing in the response functional is to multiply each quantity evaluated at the initial time by the factor $N_0$. In the remainder of the paper, only the latter requirement will be retained and the terms corresponding to the initial conditions will be systematically omitted in all of the response functionals that will be considered. \newline

Equations \ref{observable_final}--\ref{response_functional_initial_cond} form the field-theoretic version of the branching process. It presents itself under the form of a (euclidean) path integral. The further calculation of observable quantities can now be addressed thanks to the powerful machinery developed in  Quantum Field Theory (QFT) and statistical mechanics.

\section{Calculation of observables}
\label{sec:observables}

It can generally safely be assumed that in a nuclear reactor, the random power fluctuations driven by random noise are much weaker than the actual operating power level, i.e. the noise term is \textit{small}. This suggests a strategy to extract sensible results from the response functional formalism. One can expand the function $\Gamma(N)$ into the Taylor series
\begin{equation}
\label{NoisePowerSeries}
    \Gamma(N) = \lambda_0 + \lambda_1 N + \lambda_2 N^2 \, + ...,
\end{equation}
with small $\lambda_i$'s. Under this form, the noise term appearing in the response functional can be treated perturbatively, just as one does with an interaction potential in quantum mechanics. This perturbative treatment and the calculation of observables of the type given in Eq.~\ref{observable_final} is most conveniently represented under the form of Feynman diagrams \cite{greiner,zinn}.

The observables derived from the response functional will all depend on the parameters $\lambda_i$, which play here the same role as the interaction coupling constants in conventional QFT. The precise meaning of these parameters and in particular their microscopic origin is, however, left unspecified. In this respect, the response functional formalism can be viewed as a purely phenomenological approach: the precise interpretation of the coupling parameters is to be sought either a posteriori or from an external knowledge of the problem. 

One might thus rightly be worried that the calculation of observables relies upon an infinite power series expansion, whose structure and convergence depends on unknown parameters. In fact, the inclusion of powers higher than one in Eq.~\ref{NoisePowerSeries} leads in some cases to diverging expressions, rendering the perturbative expansion potentially inoperant (an example of which will be encountered in the last section). This type of divergence is a common problem in QFT and can generally be cured if one insists that all observables quantities truly describe the macro-scale behaviour of the problem, i.e. that all small scale fluctuations are truly being 'integrated over'. In a nutshell, the observables must be derived at a fixed point of a spatio-temporal scaling transformation of the system. By simple dimensional arguments \cite{janssen2004}, one finds that high order terms in the Taylor expansion of the function $\Gamma$ dies off when such a scaling operation is performed. In the Renormalization Group (RG) jargon, they represent \textit{irrelevant} parameters which formally vanish in the vicinity of the RG fixed point. These higher order terms can thus be directly discarded at this stage and $\Gamma$ may thus be simply expressed as

\begin{equation}
    \Gamma(N) =  \lambda_0 +  \lambda_1 N.
\end{equation}

\subsection{ Moments of the neutron population }

Because the noise term has zero mean, it cannot impact the mean number of neutrons $\langle N(t) \rangle = N(t)$. As long as mean values are concerned, one can thus reasonably forget about the noise term in the response functional. Using Eq.~\ref{observable_final}, the mean number of neutrons can be written as,
\begin{equation}
\label{mean_N}
    \langle N \rangle \, = \mathcal{N}  \,  \int \mathcal{D}[i\Nt] \int \mathcal{D}N \, N \, 
    e^{-  \int \left\{\Nt(t)  \left( \frac{d}{dt} - \rho\right)N(t) \right\} dt } .
\end{equation}
This expression can be simply evaluated with the help of the functional version of Wick's theorem \cite{zinn}. Taking into account for the initial condition $N(0) = N_0$, one has, 
\begin{equation}
    \langle N(t) \rangle = N_0 \times \begin{tikzpicture}[baseline=(a)]
    \begin{feynman}
      \vertex  (a) {\(t\)};
      \vertex [right=1.5cm of a](o1) {\(t_0\)};
    \diagram* {
    (a) -- (o1)
    };
    \end{feynman}
  \end{tikzpicture} =
    N_0 \, \mathcal{G}(t-t_0).
\end{equation}
$\mathcal{G}(t-t_0) = \langle N(t) \Nt(t_0) \rangle$ is the retarded Green function, solution of the equation
\begin{equation}
    \left(\frac{d}{dt} - \rho \right) \mathcal{G}(t-t') = \delta(t - t'),
\end{equation}
so that,
\begin{equation}
     \mathcal{G}(t-t') = e^{ \rho (t-t') }\, \theta(t - t'),
\end{equation}
with $\theta(t-t')$ being the Heaviside function, equals to zero whenever $t < t'$ and one otherwise. The mean number of neutrons then reads 
\begin{equation}
    \langle N(t) \rangle = N_0 \, e^{ \rho (t-t_0) }\, \theta(t - t_0).
\end{equation}

The second moment $\langle N^2(t) \rangle$ is less trivial, because it depends on the noise variance  $\Gamma(N)$. Forgetting temporarily the $\lambda_0$ term in the series expansion of $\Gamma$, and resorting again to Wick's theorem, one can symbolically write
\begin{eqnarray}
\label{observable_second_moment}
    \langle N^2(t) \rangle \, & = &  2 \lambda_1 \begin{tikzpicture}[baseline=(a)]
    \begin{feynman}[inline=(a)]
      \vertex (a);
      \vertex [above left=of a](f1);
      \vertex [below left=of a](f2);
      \vertex [right=of a](o1);
    \diagram* {
    (f1) -- (a) -- (f2),
    (a) -- (o1)
    };
    \end{feynman}
  \end{tikzpicture} 
  + 2 \; \;
  \begin{tikzpicture}[baseline=(a)]
    \begin{feynman}[inline=(a)]
      \vertex (a);
      \vertex [above= -0.3cm of a](c) ;
      \vertex [above= 0.4cm of a](b) ;
      \vertex [right=of c](o1) ;
      \vertex [right=of b](o2) ;
    \diagram* {
    (c) -- (o1),
    (b) -- (o2),
    };
    \end{feynman}
  \end{tikzpicture} \nonumber\\ & & \nonumber\\
  & = &  \langle N^2(t) \rangle_c + \langle N(t) \rangle^2,
\end{eqnarray}
where the notation $\langle \rangle_c$ is used to represent the first, so-called \textit{connected}, diagram. The variance can then be simply evaluated as ($\rho \neq 0$)
\begin{eqnarray}
  \langle N^2 \rangle - \langle N \rangle^2 & = & \langle N^2 \rangle_c  =  2 \lambda_1 N_0 \int_{t_0}^t d\tau \mathcal{G}^2(t - \tau)  \mathcal{G}(\tau - t_0) \nonumber\\
  & = &   \cfrac{ 2 \lambda_1 N_0}{\rho}  \,  e^{\rho  (t-t_0)} \left[ e^{\rho  (t-t_0)} -1  \right]. 
\end{eqnarray}
This expression exactly matches the variance one would derive from a detailed microscopic balance of the neutron population (in an infinite, homogeneous medium) \cite{paszit_pal}, provided that $\lambda_1$ be identified with the rescaled variance $(\bar{\nu}^{(2)} - \rho)/2$ of the offspring distribution of the underlying branching stochastic process. This result is a good illustration of the general outreach of the response functional formalism.  Relying only on the SDE built upon the 'coarse-grained' description of the problem, the formalism is able to take advantage of the phenomenological and microscopic random source term to extract residual information about the structure of the fluctuations around the mean field behaviour of the system. 

A strong limitation of this approach is the assumption that the noise term follows a Gaussian profile. Indeed, the calculation of higher moments $\langle N^k \rangle$ for the neutron population would formally require information about the same moments of the offspring distribution. In the response functional formalism, there is no possibility to include this type of information in the calculations. This limitation does not however applies in the present case, at least in the vicinity of criticality. It has indeed been demonstrated in \cite{garcia_millan} that the microscopic branching process under study displays a \textit{universal} behaviour in the vicinity of the critical point $\rho = 0$: it then solely depends on the reactivity $\rho$ and the second moment of its offspring distribution. Thus the Gaussian distribution imposed on the noise source does not limit the extraction of higher moments of the neutron distribution. 

The approach proposed in \cite{garcia_millan} also relies on a field-theoretic formulation of the stochastic branching process, pioneered in the work of Doi \cite{doi_1976} and Peliti \cite{peliti_1985}. Starting from a formal identification of the ladder operators $a^\dagger$ and $a$ with the multiplication or derivation by $z$ of the generating function $\psi(z,t) = \sum_n P(n,t) z^n$, one can give a structure to the stochastic process that closely resemble the structure of Fock spaces used in quantum mechanics \cite{baez_2019}. A path integral formulation of the process is then built upon the coherent states of the annihilation operator $a$ \cite{tauber_2005}.

Both the Doi-Peliti and Janssen-De Dominicis approaches allow for the derivation of a field theory, from two different starting points. In this respect, they can be viewed as complementary:
\begin{itemize}
    \item the Doi-Peliti approach relies on the explicit modelling of the underlying microscopic stochastic process. Every term in the theory is given a sound microscopical interpretation from the start. The drawback is the difficulty to include terms that cannot be described as stochastic processes (because their microscopic origin is unclear for instance);  
    
    \item the Janssen-De Dominicis approach is much more phenomenological. Its starting point is the mean field equation associated to the problem. This enables more flexibility in the modelling of phenomena that could not be given a microscopic origin (e.g. the modelling of counter-reaction mechanisms in neutronics for instance). Relying on Gaussian noise distributions and unspecified parameters, this approach is nevertheless more limited in scope when a microscopic model is available. 
\end{itemize}

Thanks to the universal behaviour of branching processes close to their critical point, both approaches coincide in the present case: they both lead to the same path integral formulation and the same diagrammatic expansion. As such, the extraction of higher moments for the neutron population in the response functional formalism exactly matches the derivation presented in \cite{garcia_millan}.  

\subsection{Addition of an external neutron source}

At low power, external neutron sources, such as the spontaneous fission on some heavy nuclei or $(\alpha,n)$ reactions might have a substantial impact on the moments of the neutron population \cite{williams_book,Dumonteil2021PatchyReactions}. The response functional formalism lends itself easily to describe this phenomenon. Considering an external neutron source, emitting an average of $S$ neutrons per second, the SDE describing the problem now reads
\begin{equation}
    \label{sde_source}
        \cfrac{d N(t)}{dt} = \rho N(t) + S + \eta(t),
\end{equation}
as well as the associated response functional,
\begin{eqnarray}
\label{action_source}
     \mathcal{S}[N,\Nt]  & = & \int \left\{\Nt(t)  \left( \frac{d}{dt} -  \rho \right)N(t) - \Nt S   \right.\nonumber\\
     & &  - \left. \Nt(t) \Gamma \Nt(t)\, \right\} dt.
\end{eqnarray}
Considering that the source term also comes with an intrinsic noise, we now can set $\lambda_0 \neq 0$ in the Taylor expansion of $\Gamma(N) = \lambda_0 + \lambda_1 N$. 

The moments of the neutron population must now take into account the external source. In the perturbative calculation, this is done in the form of new Feynman diagrams appearing in the calculations. 
For the mean number of neutrons at time $t$, one has (setting $t_0 = 0$ in the following),
\begin{eqnarray}
     \langle N(t)\rangle &=  \begin{tikzpicture}[baseline=(a)]
    \begin{feynman}[inline=(a)]
      \vertex (a);
      \vertex [right=of a](o1);
    \diagram* {
    (a) -- (o1)
    };
    \end{feynman}
  \end{tikzpicture}
  +  S \, 
  \begin{tikzpicture}[baseline=(a)]
    \begin{feynman}[inline=(a)]
      \vertex (a);
      \vertex [right=of a,crossed dot] (o1) { };
    \diagram* {
    (a) -- (o1) 
    };
    \end{feynman}
  \end{tikzpicture} \nonumber \\
  ~ \nonumber\\
  & = N_0 e^{ \rho t } + S \int_0^t e^{ \rho (t-t') } dt' \nonumber\\
  ~ \nonumber\\
  & = N_0 e^{ \rho t } + \frac{S}{\rho} \left(  e^{ \rho t } - 1 \right) ,
\end{eqnarray}
in agreement with the literature \cite{williams_book}. The second propagator that now appears in the diagram directly account for the creation of neutrons produced at a constant rate $S$ by the source. 

As for the variance of the neutron population, one has,
\begin{equation}
     \langle N^2(t)\rangle_c = 2 \lambda_1  
     \begin{tikzpicture}[baseline=(a)]
    \begin{feynman}[inline=(a)]
      \vertex (a);
      \vertex [above left=of a](f1);
      \vertex [below left=of a](f2);
      \vertex [right=of a](o1);
    \diagram* {
    (f1) -- (a) -- (f2),
    (a) -- (o1)
    };
    \end{feynman}
  \end{tikzpicture}
     +  2 \lambda_1 S
   \begin{tikzpicture}[baseline=(a)]
    \begin{feynman}[inline=(a)]
      \vertex (a);
      \vertex [above left=of a](f1);
      \vertex [below left=of a](f2);
      \vertex [right=of a, crossed dot](o1) { };
    \diagram* {
    (f1) -- (a) -- (f2),
    (a) -- (o1)
    };
    \end{feynman}
  \end{tikzpicture}
  + 2 \lambda_0
 \begin{tikzpicture}[baseline=(a)]
    \begin{feynman}[inline=(a)]
      \vertex [empty dot] (a) { };
      \vertex [above left=of a](f1);
      \vertex [below left=of a](f2);
    \diagram* {
    (f1) -- (a) -- (f2),
    };
    \end{feynman}
  \end{tikzpicture}
\end{equation}
The first diagram has already been calculated in the previous section. The following two diagrams take into account respectively:
\begin{itemize}
    \item the variance of the neutrons' branching process after their creation by the source $S$;
    \item the intrinsic variance $\lambda_0$ of the neutron source.
\end{itemize}
The calculation of the two unknown diagrams leads to the final expression ($\rho \neq 0$):
\begin{eqnarray}
 \langle N^2(t)\rangle_c & = & \frac{ 2 \lambda_1 N_0 }{\rho}  e^{\rho t } \left( e^{\rho t } - 1 \right)  + 
 \frac{2 \lambda_1}{\rho} \left( e^{\rho t } - 1 \right)^2 \nonumber\\
 & + & \frac{\lambda_0}{\rho}   \left( e^{  2 \rho t } - 1  \right).
\end{eqnarray}
 The calculation of moments of higher orders now poses a real problem. Although the universal behavior of the neutrons' branching process could, here also, be invoked, one is now limited by the noisy component of the external source term. Only if the source possesses a Gaussian distribution can the calculation be taken further and the moments $\langle N^k\rangle_c$ with $k>2$ be exactly calculated.

\section{Two species model}

Another key aspect in the modelling of the temporal behavior of the neutron population in a nuclear reactor is the role played by the precursors of delayed neutrons $M(t)$. In rare cases indeed, an induced fission does not emit all the outgoing neutrons instantaneously as the fission fragments can also emit neutrons on longer time scale controlled by their electroweak decay, ranging from a second to a few minutes, orders of magnitudes longer than prompt neutrons and occurring in the present model at a rate $\lambda_D$. Thus, while emitting a small fraction $\beta$of the overall number of neutrons, precursors control the kinetics of nuclear reactor which could not be operated on human time scales otherwise.

Including the coupling between the neutron and precursor populations, the point-like model transforms to (with $r = \rho - \beta)$ 
\begin{equation}
\label{pointKinetics}
\begin{cases}
        \cfrac{d N(t)}{dt} & = r N(t) + \lambda_D M(t) + \eta_1(t), \\
         \\
    \cfrac{d M(t)}{dt} & = \beta N(t) - \lambda_D M(t) + \eta_2(t) ,
\end{cases}
\end{equation}
where external sources have been neglected, for simpli\-city. As before, the two random source terms are supposed to follow Gaussian probability distributions with zero mean and a covariance given by the relation ($a,b = 1,2$), 
\begin{equation}
    \langle \eta_a(t) \eta_b(t') \rangle = 2  \, \Gamma_{ab}(N, M) \,  \delta(t - t'),
\end{equation}
The derivation of the response functional associated to this coupled system of equations still follows the same recipe as the one described in section \ref{sec:FieldFormulation}. Four independent fields $\Nt, \Mt, M$ and $N$ are now introduced into the theory and the response functional can be written $\mathcal{S} = \mathcal{S}_{\text{neut.}} + \mathcal{S}_{\text{prec.}} + \mathcal{S}_{\Gamma}$, with,
\begin{equation}
\label{action_neutron_2species}
   \mathcal{S}_{\text{neut.}}  =  \int \left\{\Nt(t)  \left( \frac{d}{dt}-r\right)N(t) -  \lambda_D \Nt(t) M(t)  \right\} dt,
\end{equation}

\begin{equation}
\label{action_prec_2species}
     \mathcal{S}_{\text{prec.}} =  \int \left\{ \,\Mt(t)  \left( \frac{d}{dt} + \lambda_D \right)M(t) - \beta \Mt(t) N(t) \right\} dt,
\end{equation}
and, in its most general form,
\begin{eqnarray}
       & & \mathcal{S}_{\Gamma} =  \int \Gamma_{11} \Nt^2 +  \Gamma_{22} \Mt^2 + 2 \Gamma_{12} \Nt \Mt  \nonumber\\
 & =&  \int   \left\{ \lambda_1 \Nt^2 N 
  +  \lambda_2 \Nt^2 M + \lambda_3 \Mt^2 N \right.\nonumber\\ 
   &\hspace{0.3 cm}&  + \left. \lambda_4 \Mt^2 M  + 2 \lambda_5 \Nt \Mt  N + 2 \lambda_6 \Nt \Mt M\right\} dt.
\end{eqnarray}
Two Green functions are to be associated to the free parts of the neutron and precursor functionals,
\begin{equation}
   \mathcal{G}_N(t - t') =  \begin{tikzpicture}[baseline=(a)]
    \begin{feynman}
      \vertex  (a) {\(t\)};
      \vertex [right=1.5cm of a](o1) {\(t'\)};
    \diagram* {
    (a) -- (o1)
    };
    \end{feynman}
  \end{tikzpicture} =
    e^{ r(t-t') } \theta(t - t')
\end{equation}
and
\begin{equation}
   \mathcal{G}_M(t - t') =  \begin{tikzpicture}[baseline=(a)]
    \begin{feynman}
      \vertex  (a) {\(t\)};
      \vertex [right=1.5cm of a](o1) {\(t'\)};
    \diagram* {
    (a) -- [scalar] (o1)
    };
    \end{feynman}
  \end{tikzpicture} =
    e^{ -\lambda_D (t-t') } \theta(t - t').
\end{equation}

Because of the couplings between the neutron and precursor populations appearing in Eq.~\ref{action_neutron_2species} and Eq.~\ref{action_prec_2species}, observables must however be ultimately  constructed from a generalized set of propagators, which solve the full coupled set of equations Eq.~\ref{pointKinetics}. These generalized propagators may be symbolically written, 

\begin{widetext}
\begin{eqnarray}
 &  & \Sigma_{NN}(t,t') = \begin{tikzpicture}[baseline=(a)]
    \begin{feynman}[small, inline=(a)]
      \vertex (a);
      \vertex [right=0.5 cm of a,blob](b) {};
      \vertex [right=0.7 cm of b](o1);
    \diagram* {
    (a) --  (b) -- (o1)
    };
    \end{feynman}
  \end{tikzpicture} =
     \begin{tikzpicture}[baseline=(a)]
    \begin{feynman}[inline=(a)]
      \vertex (a);
      \vertex [right=of a](o1);
    \diagram* {
    (a) -- (o1)
    };
    \end{feynman}
  \end{tikzpicture}
  + \beta \, \lambda_D \,
    \begin{tikzpicture}[baseline=(a)]
    \begin{feynman}[small, inline=(a)]
      \vertex (a);
      \vertex [right=0.5 cm of a,dot](b) {};
      \vertex [right=0.7 cm of b,dot](c) {};
      \vertex [right=0.5 cm of c](o1);
    \diagram* {
    (a) -- (b) ,
    (b)--[scalar] (c),
    (c) -- (o1)
    };
    \end{feynman}
  \end{tikzpicture}
  + \left( \beta \, \lambda_D  \right)^2
  \begin{tikzpicture}[baseline=(a)]
    \begin{feynman}[small, inline=(a)]
      \vertex (a);
      \vertex [right=0.5 cm of a,dot](b) {};
      \vertex [right=0.7 cm of b,dot](c) {};
      \vertex [right=0.7 cm of c,dot](d) {};
      \vertex [right=0.7 cm of d,dot](e) {};
      \vertex [right=0.5 cm of e](o1);
    \diagram* {
    (a) -- (b) ,
    (b)--[scalar] (c),
    (c) -- (d) ,
    (d)--[scalar] (e),
    (e) -- (o1)
    };
    \end{feynman}
  \end{tikzpicture}
  + ... \, , \, \label{sig_nn} \\ & & \nonumber \\
    & & \Sigma_{MM}(t,t')  =
    \begin{tikzpicture}[baseline=(a)]
    \begin{feynman}[small, inline=(a)]
      \vertex (a);
      \vertex [right=0.5 cm of a,blob](b) {};
      \vertex [right=0.7 cm of b](o1);
    \diagram* {
    (a) -- [scalar] (b) --[scalar] (o1)
    };
    \end{feynman}
  \end{tikzpicture} =
     \begin{tikzpicture}[baseline=(a)]
    \begin{feynman}[inline=(a)]
      \vertex (a);
      \vertex [right=of a](o1);
    \diagram* {
    (a) --[scalar] (o1)
    };
    \end{feynman}
  \end{tikzpicture}
  + \beta \, \lambda_D \,
    \begin{tikzpicture}[baseline=(a)]
    \begin{feynman}[small, inline=(a)]
      \vertex (a);
      \vertex [right=0.5 cm of a,dot](b) {};
      \vertex [right=0.7 cm of b,dot](c) {};
      \vertex [right=0.5 cm of c](o1);
    \diagram* {
    (a) --[scalar] (b) ,
    (b)-- (c),
    (c) --[scalar] (o1)
    };
    \end{feynman}
  \end{tikzpicture}
  + \left( \beta \, \lambda_D  \right)^2
  \begin{tikzpicture}[baseline=(a)]
    \begin{feynman}[small, inline=(a)]
      \vertex (a);
      \vertex [right=0.5 cm of a,dot](b) {};
      \vertex [right=0.7 cm of b,dot](c) {};
      \vertex [right=0.7 cm of c,dot](d) {};
      \vertex [right=0.7 cm of d,dot](e) {};
      \vertex [right=0.5 cm of e](o1);
    \diagram* {
    (a) --[scalar] (b) ,
    (b)-- (c),
    (c) --[scalar] (d) ,
    (d)-- (e),
    (e) --[scalar] (o1)
    };
    \end{feynman}
  \end{tikzpicture}
  + ... \, , \, \\ & & \nonumber \\
& & \Sigma_{NM}(t,t')  =  \begin{tikzpicture}[baseline=(a)]
    \begin{feynman}[small, inline=(a)]
      \vertex (a);
      \vertex [right=0.5 cm of a,blob](b) {};
      \vertex [right=0.7 cm of b](o1);
    \diagram* {
    (a) -- [scalar] (b) --(o1)
    };
    \end{feynman}
  \end{tikzpicture} = \beta \left\{ \begin{tikzpicture}[baseline=(a)]
    \begin{feynman}[inline=(a)]
      \vertex (a);
      \vertex [right=of a,dot](b) {};
      \vertex [right=of b](o1);
    \diagram* {
    (a) --[scalar] (b),
    (b) -- (o1)
    };
    \end{feynman}
  \end{tikzpicture}
  + \beta \, \lambda_D \, \, 
    \begin{tikzpicture}[baseline=(a)]
    \begin{feynman}[inline=(a)]
      \vertex (a);
      \vertex [right=of a,dot](b) {};
      \vertex [right=of b,dot](c) {};
      \vertex [right=of c,dot](d) {};
      \vertex [right=of d](o1);
    \diagram* {
    (a) --[scalar] (b) ,
    (b)-- (c),
    (c)-- [scalar] (d),
    (d) -- (o1)
    };
    \end{feynman}
  \end{tikzpicture}
  + ...\right\} , \\ & & \nonumber\\
 & &  \Sigma_{MN}(t,t')  =  \begin{tikzpicture}[baseline=(a)]
    \begin{feynman}[small, inline=(a)]
      \vertex (a);
      \vertex [right=0.5 cm of a,blob](b) {};
      \vertex [right=0.8 cm of b](o1);
    \diagram* {
    (a) --  (b) -- [scalar] (o1)
    };
    \end{feynman}
  \end{tikzpicture} = \lambda_D \left\{ \begin{tikzpicture}[baseline=(a)]
    \begin{feynman}[inline=(a)]
      \vertex (a);
      \vertex [right=of a,dot](b) {};
      \vertex [right=of b](o1);
    \diagram* {
    (a) -- (b),
    (b) --[scalar] (o1)
    };
    \end{feynman}
  \end{tikzpicture}
  + \beta \, \lambda_D \, \, 
    \begin{tikzpicture}[baseline=(a)]
    \begin{feynman}[inline=(a)]
      \vertex (a);
      \vertex [right=of a,dot](b) {};
      \vertex [right=of b,dot](c) {};
      \vertex [right=of c,dot](d) {};
      \vertex [right=of d](o1);
    \diagram* {
    (a) -- (b) ,
    (b)-- [scalar](c),
    (c)-- [] (d),
    (d) --[scalar] (o1)
    };
    \end{feynman}
  \end{tikzpicture}
 + ...\right\}.  
\end{eqnarray}
\end{widetext}
The calculation of each one of the $\Sigma_{IJ}$ involves an infinite series of nested integrals. Fortunately, these infinite series can be summed over and it is possible to obtain analytical results in all four cases (see annex \ref{app_a}). 
A much more convenient but indirect way to extract an analytical expression out of these propagators is to realize that they all directly correspond to the mean value of either the neutron or precursor populations for different initial conditions. Indeed, setting $( N(t_0), M(t_0)) =(N_0, 0 )$, one can show that,
    \begin{equation}
        \langle N(t) \rangle_{(N_0, 0 )} = N_0 \times \Sigma_{NN}(t,t_0),
    \end{equation}
    \begin{equation}
        \langle M(t) \rangle_{(N_0, 0 )} = N_0 \times \Sigma_{NM}(t,t_0),
    \end{equation}
whereas, for $( N(t_0), M(t_0)) =(0, M_0 )$, 
    \begin{equation}
        \langle N(t) \rangle_{(0, M_0 )} = M_0 \times \Sigma_{MN}(t,t_0),
    \end{equation}
    \begin{equation}
        \langle M(t) \rangle_{(0, M_0 )} = M_0 \times \Sigma_{MM}(t,t_0),
    \end{equation}
so that each of these propagators is the solution of the system of equations Eq.~\ref{pointKinetics} for a specific set of initial conditions (and with the stochastic terms set to zero). This system of equations can be straightforwardly integrated and solutions extracted. The explicit form of each generalized propagator is provided in annex \ref{app_a}.

Equipped with an analytical expression for the propagators of the theory, the variance of the process may then be evaluated. Imposing the initial condition $( N(0), M(0)) =(N_0, 0 )$  , the variance of the neutron and precursor populations may be diagrammatically written,  
\begin{widetext}
\begin{eqnarray}
\label{var_N}
     \langle N^2(t) \rangle_c & = & 2 N_0 \lambda_1 \begin{tikzpicture}[baseline=(a)]
    \begin{feynman}[inline=(a)]
      \vertex (a);
      \vertex [right= 0.5 cm of a,blob](b) {};
      \vertex [above left=0.5 cm of a,blob](f1) {};
      \vertex [below left=0.5cm of a,blob](f2) {};
      \vertex [above left=0.7 cm of f1](ff1);
      \vertex [below left=0.7 cm of f2](ff2);
      \vertex [right=0.7 cm of b](o1);
    \diagram* {
     (ff1) -- (f1)  -- (a) --  (f2) -- (ff2),
    (a) -- (b) --  (o1)
    };
    \end{feynman}
  \end{tikzpicture}
  + 2 N_0 \lambda_2 
  \begin{tikzpicture}[baseline=(a)]
    \begin{feynman}[inline=(a)]
      \vertex (a);
      \vertex [right= 0.5 cm of a,blob](b) {};
      \vertex [above left=0.5 cm of a,blob](f1) {};
      \vertex [below left=0.5cm of a,blob](f2) {};
      \vertex [above left=0.7 cm of f1](ff1);
      \vertex [below left=0.7 cm of f2](ff2);
      \vertex [right=0.7 cm of b](o1);
    \diagram* {
     (ff1) -- (f1) -- (a) -- (f2) -- (ff2),
    (a) -- [scalar] (b) --  (o1)
    };
    \end{feynman}
  \end{tikzpicture}
  + 2 N_0 \lambda_3
   \begin{tikzpicture}[baseline=(a)]
    \begin{feynman}[inline=(a)]
      \vertex (a);
      \vertex [right= 0.5 cm of a,blob](b) {};
      \vertex [above left=0.5 cm of a,blob](f1) {};
      \vertex [below left=0.5cm of a,blob](f2) {};
      \vertex [above left=0.7 cm of f1](ff1);
      \vertex [below left=0.7 cm of f2](ff2);
      \vertex [right=0.7 cm of b](o1);
    \diagram* {
     (ff1) -- (f1) --[scalar] (a) --[scalar]  (f2) -- (ff2),
    (a) -- (b) --  (o1)
    };
    \end{feynman}
  \end{tikzpicture} \nonumber\\
   & & + 2 N_0 \lambda_4
   \begin{tikzpicture}[baseline=(a)]
    \begin{feynman}[inline=(a)]
      \vertex (a);
      \vertex [right= 0.5 cm of a,blob](b) {};
      \vertex [above left=0.5 cm of a,blob](f1) {};
      \vertex [below left=0.5cm of a,blob](f2) {};
      \vertex [above left=0.7 cm of f1](ff1);
      \vertex [below left=0.7 cm of f2](ff2);
      \vertex [right=0.7 cm of b](o1);
    \diagram* {
     (ff1) -- (f1) --[scalar] (a) --[scalar]  (f2) -- (ff2),
    (a) --[scalar] (b) --  (o1)
    };
    \end{feynman}
  \end{tikzpicture} 
  + 2 N_0 \lambda_5
   \begin{tikzpicture}[baseline=(a)]
    \begin{feynman}[inline=(a)]
      \vertex (a);
      \vertex [right= 0.5 cm of a,blob](b) {};
      \vertex [above left=0.5 cm of a,blob](f1) {};
      \vertex [below left=0.5cm of a,blob](f2) {};
      \vertex [above left=0.7 cm of f1](ff1);
      \vertex [below left=0.7 cm of f2](ff2);
      \vertex [right=0.7 cm of b](o1);
    \diagram* {
     (ff1) -- (f1) --[scalar] (a) --  (f2) -- (ff2),
    (a) -- (b) --  (o1)
    };
    \end{feynman}
  \end{tikzpicture} 
    + 2 N_0 \lambda_6
   \begin{tikzpicture}[baseline=(a)]
    \begin{feynman}[inline=(a)]
      \vertex (a);
      \vertex [right= 0.5 cm of a,blob](b) {};
      \vertex [above left=0.5 cm of a,blob](f1) {};
      \vertex [below left=0.5cm of a,blob](f2) {};
      \vertex [above left=0.7 cm of f1](ff1);
      \vertex [below left=0.7 cm of f2](ff2);
      \vertex [right=0.7 cm of b](o1);
    \diagram* {
     (ff1) -- (f1) --[scalar] (a) --  (f2) -- (ff2),
    (a) -- [scalar] (b) --  (o1)
    };
    \end{feynman}
  \end{tikzpicture} ,
\end{eqnarray}
\begin{eqnarray}
\label{var_M}
     \langle M^2(t) \rangle_c &=& 2 N_0 \lambda_1 \begin{tikzpicture}[baseline=(a)]
    \begin{feynman}[inline=(a)]
      \vertex (a);
      \vertex [right= 0.5 cm of a,blob](b) {};
      \vertex [above left=0.5 cm of a,blob](f1) {};
      \vertex [below left=0.5cm of a,blob](f2) {};
      \vertex [above left=0.7 cm of f1](ff1);
      \vertex [below left=0.7 cm of f2](ff2);
      \vertex [right=0.7 cm of b](o1);
    \diagram* {
     (ff1) --[scalar] (f1)  -- (a) --  (f2) --[scalar] (ff2),
    (a) -- (b) --  (o1)
    };
    \end{feynman}
  \end{tikzpicture}
  + 2 N_0 \lambda_2 
  \begin{tikzpicture}[baseline=(a)]
    \begin{feynman}[inline=(a)]
      \vertex (a);
      \vertex [right= 0.5 cm of a,blob](b) {};
      \vertex [above left=0.5 cm of a,blob](f1) {};
      \vertex [below left=0.5cm of a,blob](f2) {};
      \vertex [above left=0.7 cm of f1](ff1);
      \vertex [below left=0.7 cm of f2](ff2);
      \vertex [right=0.7 cm of b](o1);
    \diagram* {
     (ff1) --[scalar] (f1) -- (a) -- (f2) --[scalar] (ff2),
    (a) -- [scalar] (b) --  (o1)
    };
    \end{feynman}
  \end{tikzpicture}
  + 2 N_0 \lambda_3
   \begin{tikzpicture}[baseline=(a)]
    \begin{feynman}[inline=(a)]
      \vertex (a);
      \vertex [right= 0.5 cm of a,blob](b) {};
      \vertex [above left=0.5 cm of a,blob](f1) {};
      \vertex [below left=0.5cm of a,blob](f2) {};
      \vertex [above left=0.7 cm of f1](ff1);
      \vertex [below left=0.7 cm of f2](ff2);
      \vertex [right=0.7 cm of b](o1);
    \diagram* {
     (ff1) --[scalar] (f1) --[scalar] (a) --[scalar]  (f2) --[scalar] (ff2),
    (a) -- (b) --  (o1)
    };
    \end{feynman}
  \end{tikzpicture} \nonumber \\
   & & + 2 N_0 \lambda_4
   \begin{tikzpicture}[baseline=(a)]
    \begin{feynman}[inline=(a)]
      \vertex (a);
      \vertex [right= 0.5 cm of a,blob](b) {};
      \vertex [above left=0.5 cm of a,blob](f1) {};
      \vertex [below left=0.5cm of a,blob](f2) {};
      \vertex [above left=0.7 cm of f1](ff1);
      \vertex [below left=0.7 cm of f2](ff2);
      \vertex [right=0.7 cm of b](o1);
    \diagram* {
     (ff1) --[scalar] (f1) --[scalar] (a) --[scalar]  (f2) --[scalar] (ff2),
    (a) --[scalar] (b) --  (o1)
    };
    \end{feynman} 
  \end{tikzpicture} 
    + 2 N_0 \lambda_5
   \begin{tikzpicture}[baseline=(a)]
    \begin{feynman}[inline=(a)]
      \vertex (a);
      \vertex [right= 0.5 cm of a,blob](b) {};
      \vertex [above left=0.5 cm of a,blob](f1) {};
      \vertex [below left=0.5cm of a,blob](f2) {};
      \vertex [above left=0.7 cm of f1](ff1);
      \vertex [below left=0.7 cm of f2](ff2);
      \vertex [right=0.7 cm of b](o1);
    \diagram* {
     (ff1) --[scalar] (f1) --[scalar] (a) --  (f2) --[scalar](ff2),
    (a) -- (b) --  (o1)
    };
    \end{feynman}
  \end{tikzpicture} 
    + 2 N_0 \lambda_6
   \begin{tikzpicture}[baseline=(a)]
    \begin{feynman}[inline=(a)]
      \vertex (a);
      \vertex [right= 0.5 cm of a,blob](b) {};
      \vertex [above left=0.5 cm of a,blob](f1) {};
      \vertex [below left=0.5cm of a,blob](f2) {};
      \vertex [above left=0.7 cm of f1](ff1);
      \vertex [below left=0.7 cm of f2](ff2);
      \vertex [right=0.7 cm of b](o1);
    \diagram* {
     (ff1) --[scalar] (f1) --[scalar] (a) --  (f2) --[scalar] (ff2),
    (a) -- [scalar] (b) --  (o1)
    };
    \end{feynman}
  \end{tikzpicture} .
\end{eqnarray}
\end{widetext}

Just as it was the case in the last section, both expressions depend on the  set of parameters $\lambda_i$, which are not specified by the theory.
With a total of 6 unknown parameters for only two observables, it may seem at first sight difficult to give a precise meaning to each term. However, the structure of the diagrammatic expansion and the results of the one species model derived earlier allows to do so, at least partially. 
In light of the results obtained in the previous section, it might indeed  be inferred that the diagrams associated to $\lambda_1$ appearing in both expressions represent the variance associated to the neutron’s offspring distribution. There is furthermore no reason to expect that the value for $\lambda_1$ obtained in the case of the one species model should be any different in the present situation,  except for the replacement of the reactivity $\rho$  with the prompt reactivity $r = \rho - \beta$.

On the same model, the diagrams associated with $\lambda_3$ are most certainly to be  identified with the variance of the precursor emitted by fission. 

The diagrams associated to $\lambda_2$ and $\lambda_4$ share the same output structure as for $\lambda_1$ and $\lambda_3$ but possess a precursor at the origin of the branching term. This suggest that these diagrams encode for a source of variance intrinsic to the precursor population. The only process responsible for such kind of fluctuation in the precursor population is their decay process which, being a Poisson process, has a variance $\lambda_D$.\newline

\begin{figure}[!ht]
\centering
\hspace{-1 cm}
\includegraphics[scale=0.75]{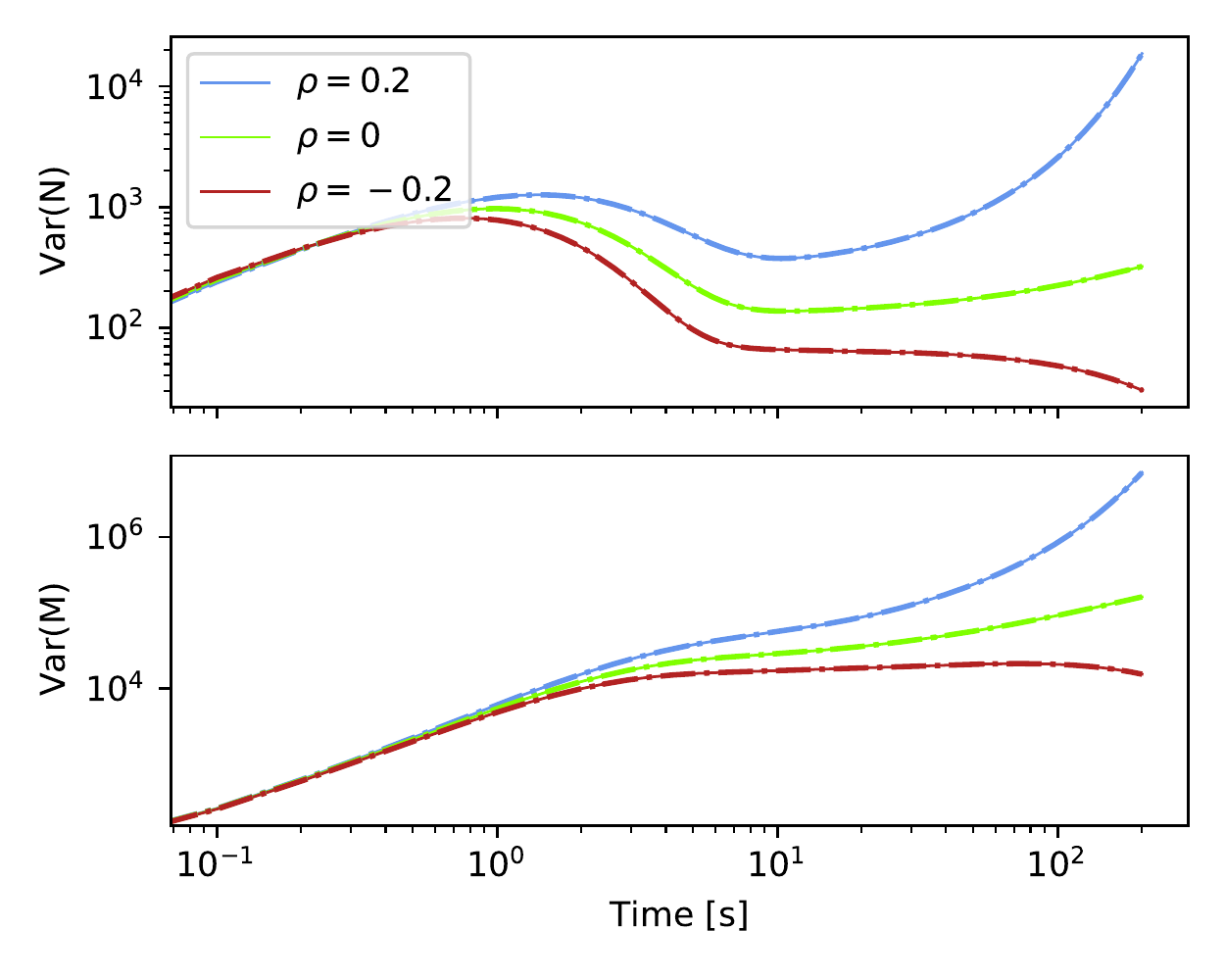}
\caption{Comparison of the variance of the neutron and precursor populations extracted from the solution of the stochastic process (plain lines) and from the calculation of Feynman diagrams (dotted lines), for $\rho =$ -0.2 (red), 0 (green), 0.2 (blue). The calculations are made assuming the following set of parameters: $\beta = 0.8$, $\lambda_D = 0.03$, $N_0 = 1000$, $\bar{\nu}_n^{(2)} = 2$, $\bar{\nu}_m^{(2)} = 1.2$. The two offspring distributions were supposed independent so that $ \overline{\nu_m(\nu_n - 1)} = \frac{\beta}{1-\beta} \bar{\nu}_n^{(2)} $. }
\label{Var_NM}
\end{figure}

At this point, one can thus postulate,
\begin{equation}
    \begin{cases}
            \lambda_1 = (\bar{\nu}_n^{(2)} - \rho + \beta)/2, \\
            \lambda_3 = (\bar{\nu}_m^{(2)} + \beta)/2, \\
            \lambda_2 = \lambda_4 = \lambda_D/2,
    \end{cases}
\end{equation}
so that only two unknown parameters, $\lambda_5$ and $\lambda_6$, remain to be determined for the two variances. Comparing equations Eq.~\ref{var_N}-\ref{var_M} with a microscopic balance of the neutron and precursor populations \cite{paszit_pal} for different values of $\rho$, $\beta$ and $\lambda_D$, one gets (the offspring distributions of neutrons and precursors were supposed independent), 
\begin{equation}
    \begin{cases}
            \lambda_5 =  \overline{\nu_m ( \nu_n -1)} = \bar{\nu}_n^{(2)} \frac{\beta}{1-\beta}, \\
            \lambda_6 = - \lambda_D.
    \end{cases}
\end{equation}
A comparison of the diagrammatic expansion with the results of the literature is presented in Figure \ref{Var_NM}. The temporal evolution of the variance of both the neutron and precursor populations are displayed, in all three regimes: sub-critical ($\rho<0)$, critical $(\rho=0)$ and super-critical $(\rho>0)$. An excellent agreement is found in all three cases.

\section{Stochastic diffusion equation}

Now equipped with the main phenomena of 0 dimensional neutronics, our field theoretic model can transition to proper neutron transport theory. A convenient and easy way to do so, yet realistic \footnote{the diffusion equation accurately reproduces the transport equation for isotropic media and far from the boundaries, see \cite{Duderstadt1979TransportTheory} for more details} (it is used in the operational context of reactor core monitoring), is to lean on the Brownian motion and its associated diffusion equation. The diffusion equation, which can also formally be built as a coarse grained description of a microscopic, stochastic process. When supplemented by an external random noise term, the (one energy group) equation can be written as
\begin{equation}
    \label{stoch_diffusion}
    \frac{\partial }{ \partial t} N(\vec{x},t) - D ( \Delta + r)  N(\vec{x},t) + \eta(\vec{x},t) = 0,
\end{equation}
where the rescaled parameter $r = \rho/D$ as been defined for convenience and where now $N(\vec{x},t)$ should be interpreted as the neutron density field. The transcription of this $d+1$ diffusion equation into a response functional follows, here also, exactly the same line as in section \ref{sec:FieldFormulation}. One thus have (in $d$ spatial dimensions), 
\begin{equation}
    \mathcal{S} = \int d^d \vec{x} \int dt \Nt(\vec{x},t) \left( \partial_t - D ( \Delta + r)  \right) N(\vec{x},t) - \Nt \Gamma \Nt.
\end{equation}
The propagator of the theory is most conveniently written in the Fourier transformed spatial domain ($\vec{p}$,$t$), 
 \begin{equation}
    \mathcal{G}(\vec{p},t) = e^{ -D(\vec{p}^{\,2} - r)t }.
\end{equation}
The structure of the interaction term of the response functional is not affected by the dimensionality of the problem. As a consequence, the diagrammatic expansion of the theory is the same as the one obtained in section \ref{sec:observables}. It can for instance be immediately concluded that provided with the initial condition $ N(\vec{x},t) = N_0 \, \delta(t) \delta^{(d)}(\vec{x})$, one finds that
\begin{equation}
    \langle N(\vec{x},t) \rangle = N_0  \mathcal{G}(\vec{x},t) = N_0 \, \sqrt{\frac{\pi}{4Dt}} \,  e^{ r D t - \frac{|\vec{x}|^2}{4 D t} }.
\end{equation}
A substantial amount of work has been done in the past, revolving around the possibility that neutrons evolving in fissile matter might gather into clusters and that the neutron density map should consequently present patchiness. This neutron clustering phenomenon results from the very nature of the underlying branching process they are subjected to \cite{Dumonteil2014ParticleSimulations, Zoia2014ClusteringGeometries,  DeMulatier2015TheRevisited,Agarwal2001, Sutton2017NeutronCalculations,Dumonteil2017ClusteringMedia,Dumonteil2021PatchyReactions}, and more specifically from the asymmetry between their uniform capture in the medium and their production only at spatial sites correlated to their parents \. This phenomenon can be quantified through the equal-time spatial correlation function $\langle N(\vec{x},t) N(\vec{y},t) \rangle$ which, in the response functional formal\-ism can be written as
\begin{widetext}
\begin{align}
\label{spatial_corr_func}
    & \langle N(\vec{x},t) N(\vec{y},t) \rangle_c  = 2 \lambda_1
    \begin{tikzpicture}[baseline=(a)]
    \begin{feynman}[inline=(a)]
      \vertex (a);
      \vertex [above left=of a](f1);
      \vertex [below left=of a](f2);
      \vertex [right=of a](o1);
    \diagram* {
     (f1)  -- [anti fermion] (a) -- [fermion] (f2),
    (a) -- [anti fermion] (o1)
    };
    \end{feynman}
  \end{tikzpicture}
    = 2 \lambda_1 \int d^d\vec{w}  \int_0^t dt_1  \mathcal{G}(\vec{w},t_1) \mathcal{G}(\vec{x}-\vec{w},t-t_1) \mathcal{G}(\vec{y}-\vec{w},t'-t_1) \nonumber\\
    & = 2 \lambda_1 \int_0^t dt_1 \iiint  \frac{d^d\vec{p_1}}{(2\pi)^d} \frac{d^d\vec{p_2}}{(2\pi)^d} \frac{d^d\vec{p_3}}{(2\pi)^d} \mathcal{G}(\vec{p_1},t_1) \mathcal{G}(\vec{p_2},t-t_1) \mathcal{G}(\vec{p_3},t-t_1) e^{-i \vec{p_2} \cdot \vec{x}} e^{-i \vec{p_3} \cdot \vec{y}} \int d^d\vec{w} e^{-i \vec{w} \cdot (\vec{p_1} - \vec{p_2} - \vec{p_3})} \nonumber\\
    & = 2 \lambda_1 \int_0^t dt_1 \iiint  \frac{d^d\vec{p_1}}{(2\pi)^d} \frac{d^d\vec{p_2}}{(2\pi)^d} \frac{d^d\vec{p_3}}{(2\pi)^d} \mathcal{G}(\vec{p_1},t_1) \mathcal{G}(\vec{p_2},t-t_1) \mathcal{G}(\vec{p_3},t-t_1) e^{-i \vec{p_2} \cdot \vec{x}} e^{-i \vec{p_3} \cdot \vec{y}} (2\pi)^d \delta^{(d)}( \vec{p_1} - \vec{p_2} - \vec{p_3} ) \nonumber\\
    & = 2 \lambda_1 \int_0^t dt_1 \iint  \frac{d^d\vec{p_2}}{(2\pi)^d} \frac{d^d\vec{p_3}}{(2\pi)^d} \mathcal{G}(\vec{p_2} + \vec{p_3},t_1) \mathcal{G}(\vec{p_2},t-t_1) \mathcal{G}(\vec{p_3},t-t_1) e^{-i \vec{p_2} \cdot \vec{x}} e^{-i \vec{p_3} \cdot \vec{y}}.
\end{align}
 \end{widetext}
Note that because of the appearance of Dirac delta functions enforcing momentum conservation at each vertex, it becomes useful to indicate on the Feynman graphs the directions of the flowing momenta. 

To proceed further with the calculation, one has to pay special attention to the initial condition chosen for the neutron population. In the case of neutrons evolving in an homogeneous and infinite system, and setting a uniform initial neutron density $N_0$, one can use the translation invariance of the problem to set $\vec{p_1} = \vec{0}$. Momentum conservation then implies that Eq.~\ref{spatial_corr_func} be transformed into
\begin{eqnarray}
    \langle N(\vec{x},t) N(\vec{y},t) & &\rangle_c  =  2  \lambda_1 N_0 
     \int_0^t dt_1 \int  \frac{d^d\vec{p}}{(2\pi)^d} \mathcal{G}(\vec{0},t_1) \times \nonumber\\
    & &\mathcal{G}(\vec{p},t-t_1) \mathcal{G}(-\vec{p},t-t_1) e^{-i \vec{p} \cdot (\vec{x} - \vec{y})}.
\end{eqnarray}
Performing the integration over the momentum $\vec{p}$ in $d$-dimension, one obtains,
\begin{equation}
    \langle N(\vec{x},t) N(\vec{y},t) \rangle_c  = 2 \lambda_1 N_0 e^{D r t} \int_0^t  \frac{e^{Drt'}}{(8\pi D t')^{d/2}} \, e^{ - \frac{|\vec{x} - \vec{y}|^2}{ 8 D t'} } dt' ,
\end{equation}
in agreement with the literature \cite{mulatier2015}.

\section{From neutronics to reactor physics - addition of a simple feedback mechanism }
\label{sec:directed_percolation}

Reactor physics is neutron transport theory supplemented by feedback mechanisms, which happen whenever the reactor is operated at high power. 
These feedback mechanisms are a central aspect of the stability of nuclear reactors but their detailed modelling is generally difficult to implement in a neutron balance, principally because they cannot be associated to a microscopic stochastic process: for example, the Doppler effect is proportional to local neutron densities, since neutrons heat up their surrounding media. In the response functional formalism, these effect can however be directly incorporated at the macroscopic level of description, i.e. in the mean field equations. 


In the following, this advantage is therefore exploited to explore a simplified but realistic model of a reactor in the presence of counter-reactions. It will be shown that the feedback mechanism imposed on the neutron population has a strong impact on the structure of the fluctuations close to the percolation transition. The field theoretic viewpoint adopted here will in particular allow to establish that the neutron gas undergoes, whenever the reactor is exactly critical, a phase transition whose universality class will be identified.

\subsection{A simplified model with Doppler broadening}

One of the most paradigmatic feedback effect affecting the neutron population in a reactor core is the Doppler broadening of the capture resonances of $^{238}$U, which couples the neutron population to the temperature of the fuel. As the temperature increases, more and more neutrons are captured by the medium and disappear: the overall neutron population is pulled back towards a reference operating power. This feedback is quite difficult to include in a detailed microscopic model because it involves the reactor power or the temperature of the medium which are macroscopic quantities. It can however be straightforwardly implemented under the form of a restoring force of the type $\sigma N(N - N_{\text{ref.}})$ in the diffusion equation \cite{bell_glasstone}. The behavior of the neutron population evolving in a nuclear reactor operating at nominal power (and thus subject to Doppler broadening) can consequently be described by the equation 
\begin{equation}
    \label{stoch_diffusion_doppler}
    \frac{\partial }{ \partial t} N(\vec{x},t) - D ( \Delta + r)  N(\vec{x},t)  +  \sigma N^2(\vec{x},t)  = 0,
\end{equation}
where the shifted parameter $ r = (\rho + \sigma N_{\text{ref.}})/D$ has been introduced, for convenience.

The response functional and the resulting theory to which this equation is associated has been extensively studied. It corresponds to the functional of the \textit{Directed Percolation} (DP) universality class \cite{janssen2004,tauber_2005}:
\begin{equation}
    \mathcal{S} = \int d^d \vec{x} \int dt \Nt \left( \partial_t - D ( \Delta + r)  \right) N + \sigma \Nt N^2 - \lambda \Nt^2 N  .
\end{equation}

The inclusion of the term $\sigma N^2$ has a strong impact on the structure of the theory, for now the two following processes are allowed
\begin{equation*}
    2 \lambda \begin{tikzpicture}[baseline=(a)]
    \begin{feynman}[inline=(a)]
      \vertex (a);
      \vertex [above left=of a](f1);
      \vertex [below left=of a](f2);
      \vertex [right=of a](o1);
    \diagram* {
    (f1) --[anti fermion] (a) --[fermion] (f2),
    (a) --[anti fermion] (o1)
    };
    \end{feynman}
  \end{tikzpicture} \hspace{2 cm} 2 \sigma
      \begin{tikzpicture}[baseline=(a)]
    \begin{feynman}[inline=(a)]
      \vertex (a);
      \vertex [above right=of a](f1);
      \vertex [below right=of a](f2);
      \vertex [left=of a](o1);
    \diagram* {
    (f1) --[fermion] (a) --[anti fermion] (f2),
    (a) --[fermion] (o1)
    };
    \end{feynman}
  \end{tikzpicture}
\end{equation*}
as well as their combination, i.e. the loop diagram
\begin{equation}
    4 \sigma \lambda \, 
    \begin{tikzpicture}[baseline=(a)]
    \begin{feynman}[inline=(a)]
      \vertex (z);
      \vertex [right=of z](a);
      \vertex [right=of a] (b);
      \vertex [above left=of a](f1);
      \vertex [below left=of a](f2);
      \vertex [right=of b](o1);
    \diagram* {
     (z) --[anti fermion] (a) 
     -- [fermion,half left] (b)
     -- [fermion,half left] (a),
    (b) --[anti fermion] (o1)
    };
    \end{feynman}
  \end{tikzpicture} \, ,
\end{equation}
whose contribution leads to badly divergent behaviors of two different types \cite{tauber_2005}:
\begin{itemize}
    \item a divergence is observed in dimension $d \geq 2$, because the momentum  associated to the loop is not constrained by any conservation rule. Its integration thus runs over theoretically infinite momenta and include small distances, short time, unbounded fluctuations of the neutron population. This is a paradigmatic example of an Ultra-Violet (UV) divergence;  
    
    \item another and more serious divergence occurs in dimension $d \leq 4$. An Infrared (IR) divergence is observed in the long time, long distance limit. The highest dimension $d_c = 4$ for which the divergence is observed is called the \textit{critical dimension} of the problem.   
\end{itemize}

Both types of divergences can be eliminated through the well-known renormalization procedure \cite{Delamotte_2004}, which, in essence, consists in the application of successive spatio-temporal scaling transformations on the problem, so that the small distance and UV divergent fluctuations be truly integrated over. The true macro-scale description of the problem is obtained at a fixed point of the iterated scaling process which generally has the induced benefit of taming the residual IR divergence of the problem. 

Although the starting point of the model are the mean field equations of the problem, the renormalized observables one ultimately extract from the theory present asymptotic behaviors (i.e. critical exponents) that differ from the predictions of the mean field equations, at least in the vicinity of the phase transition. Close to the critical point, mean field theory breaks down and only the renormalized field theoretic description of the system can correctly describe the evolution of observables. \newline

\subsection{The directed percolation transition}

The largest deviation from the mean field predictions is obtained in the 0D model. Indeed, setting $D =0$ in Eq.~\ref{stoch_diffusion_doppler}, the point-like DP model can be defined as
\begin{equation}
    \frac{dN}{dt} = r N - \sigma N^2 .
\end{equation}
As long as $r>0$, this equation possesses a positive stationary solution $N_{\infty} \propto r^\beta$ ($\beta = 1$ is the first critical exponent of directed percolation). A detailed analysis with the help of the RG leads to a strikingly different prediction \cite{Canet2006,Benitez2013}: in zero dimension, whatever the values chosen for $r$ and $\sigma$, the neutron population always asymptotically vanish. This finding is in sharp contrast with the solution of the mean field equations but has been tested and validated against Monte-Carlo simulations. 
Translated into the domain of nuclear reactor physics, this means that adding a feedback term of the form $\sigma N^2$ into the point kinetics equation leads to a theory that do not possess a stable, non-trivial, equilibrium solution:  every point-like reactors subject to counter-reactions are sub-critical~!

The triviality of this point-like model can be easily circumvented though, by considering instead the case of a tri-dimensional, infinite and homogeneous reactor. The system then do admit $N_{\infty} \propto r^\beta$ as a solution, but the RG predicts a value for the exponent $\beta$ close to 0.8 in 3$d$ \cite{tauber_2005, janssen2004} (as compared to 1 for mean field). Although conceptually important, this deviation in the scaling relation between the power level and the reactivity of the reactor would be difficult to observe in practice.  \newline

Even though the feedback mechanism proposed in this paper might appear as a simplistic approximation of the true processes at play in an operating reactor, the asymptotic scaling laws and the overall behaviors predicted by the RG analysis of the directed percolation process might still persist in a more refined treatment. The \textit{DP conjecture} \cite{Hinrichsen2000,janssen2004} indeed states that the DP universality class is the most general universality class observed in transitions between an active to an absorbing state and persist under very general conditions. The modelling of critical nuclear reactors might thus fall into the general class of the DP percolation transitions, even with a more elaborate modelling of counter-reaction mechanisms.

Observable consequences of the DP transition and direct violation of the mean field equation predictions might however be difficult to observe in practice. Apart from the exponent $\beta$, two other critical exponents characterize the DP universality class. They are associated to the correlation length $\xi$ and the diffusion time scale $\tau$ that naturally appear in equation \ref{stoch_diffusion_doppler}:
\begin{equation}
\label{scalinglaws}
    \xi \sim |r|^{-\nu} \; \; \text{ and } \; \; \tau \sim \xi^z/D \sim |r|^{-z \nu}.
\end{equation}
The DP class is entirely characterized by these exponents and any observable consequence of the DP transition can be traced back to a deviation of these exponents from their mean field predictions. Deviations from the mean field theory are expected to be larger for system in lower dimensions, in models farther away from the critical dimension $d_c=4$. Table \ref{critical_exponent_DP} list the values of the exponents of directed percolation, extracted from numerical simulations, for systems in 1, 2 or 3 dimensions. 
\begin{center}
\begin{table}[h!]
\caption{Critical exponents of the DP phase transition \cite{Hinrichsen2008}. \label{critical_exponent_DP} }
\begin{tabular}{|c | c | c| c | c|} 
 \hline
 exponent & d = 1  & d = 2 & d = 3 & mean field \\
 \hline\hline
 $\beta$ & 0.276486(8) & 0.5834(30)  & 0.813(9) & 1  \\ 
 \hline
 $\nu$ & 1.096854(4) & 0.7333(75) & 0.584(5) & 0.5  \\
 \hline
 $z$ & 1.580745(10) & 1.7660(16) & 1.901(5) & 2  \\
 \hline
\end{tabular}
\end{table}
\end{center}

A detailed account of all of the possible experimental evidences of the DP transition in nuclear reactors would by far exceed the scope of this paper. As a matter of illustration, the following section will describe what observable and concrete consequence the dynamic exponent $z$ might have in a real reactor. 

\subsection{A Consequence on reactor physics}

Reactor physics aims at accurately calculating (with a precision of $\sim 1\%$) the spatial distribution of the neutron flux, from which diverse measurable local quantities can be extracted, such as fission rates, neutron counts, etc. Those quantities are measured either by in-core detection devices (e.g. the aeroball measurement system) or by ex-core detection devices (using source-range, intermediate-range or power-range neutron flux instrumentations). Any significant deviation from the flux measured by these detectors to 'expected' values should trigger various safety warnings notified by the automatic protection system, ultimately leading to the core tripping procedure. Noticeably these expected values of the flux are set by an online solving of a simplified -yet robust- neutron transport model. 

\begin{figure}[!ht]
\centering
\hspace{-1 cm}
\includegraphics[scale=0.4]{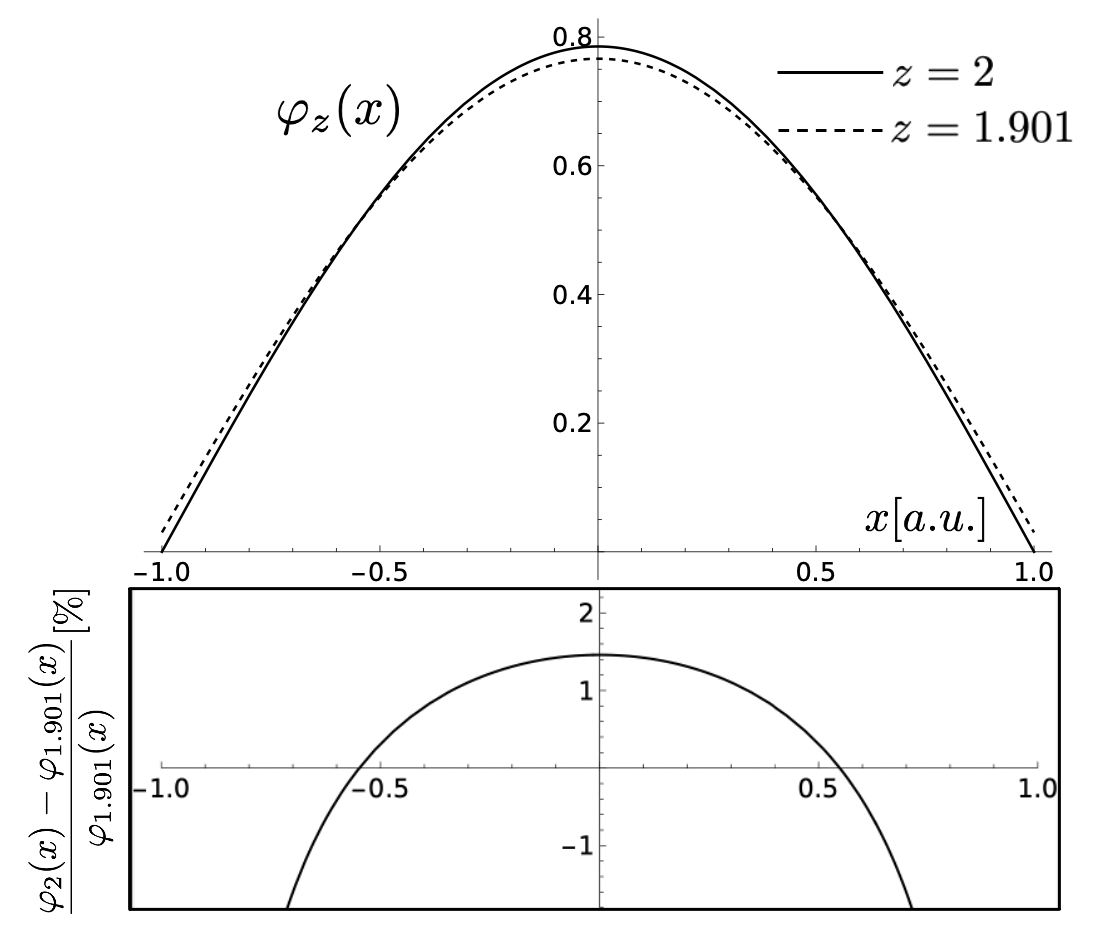}
\caption{ Top plot: comparison between the first normalized eigenfunction of the linear diffusion equation ($z=2$ in plain line) with Dirichlet boundary conditions in $-1$ and $1$, and an approximation of the first normalized eigenfunction of the fractional Laplacian (as given by \cite{Kwasnicki2012EigenvaluesInterval}) with $z=1.901$ (dashed line). The approximation is valid far from the boundaries, i.e. within the bulk. Bottom plot: relative differences (in \%) between both eigenfunctions.}
\label{compa_z}
\end{figure}

During the intermediate and power range operation of the reactor, the thermal feedback effects have to be taken into account as they modify the cross-sections dictating the laws of propagation of the neutrons. As seen in the previous section, this modification has another consequence: the macroscopic properties of the medium emerge from an intimate coupling at the mesoscopic scale between the random walk of the neutrons and the feedback reactions depending on the neutrons local densities. Usual coupling schemes relying on mean field equations are therefore not valid anymore since the mean field equations themselves are wrong, and the properties of the system can only be obtained through the critical exponents of the directed percolation defined in the previous section. In the following, since we are interested in a simplified model of reactor physics, we will restrict ourselves to $d=3$. In this case, the value of $z$, the dynamic critical exponent, is about $1.9$. This exponent describes how fast any local perturbation spreads \cite{Tauber2012CriticalBehavior} and can be related to a modification of the diffusive ($z=2$) random walk which becomes slightly super-diffusive for $d=3$ (indeed $z$ can be interpreted as an anisotropy parameter). The scaling of the spatial correlation length versus the time correlation length follows $\xi^z \sim D \tau$ (see Eq.~\ref{scalinglaws}, with $z=1.901$ in the following). These Lévy flights (see for instance \cite{Lischke2020WhatResults} for a review on fractional Laplacians and their connection to Lévy flights) call for a more quantitative modelling of the mean-field Eq.~\ref{stoch_diffusion_doppler} which is hence modified according to: 
\begin{equation}
    \label{stoch_diffusion_doppler_z}
     D (-(-\Delta)^{z/2} + r)  N(x,t)  -  \sigma N^2(x,t)  = \frac{\partial }{ \partial t} N(x,t),
\end{equation}
where $-(-\Delta)^{z/2}$ is the fractional Laplacian defined through $-(-\Delta)^{z/2}=\frac{\partial^z}{\partial |x|^z}$ ($\frac{\partial^z}{\partial |x|^z}$ being the Riesz-Feller derivative of fractional order $z$).
Yet, whenever the reactor is in a stationary state, the Doppler effect does hardly comes into action: it mainly stabilizes the core upon local perturbations or control long-term statistical fluctuations \cite{Dumonteil2021PatchyReactions}. Since the critical exponents obtained in the previous section did not explicitly depend on $\sigma$, representing the strength of the coupling between the neutron scalar field and the feedback mechanism, this equation can further be simplified by considering the parametric regime where this coupling is weak. The feedback term $\sigma N^2$ in the mean field equation can consequently be safely neglected. In this case the stationary behavior of Eq.~\ref{stoch_diffusion_doppler_z} boils down to
\begin{equation}
    \label{stoch_diffusion_doppler_nofeedback}
     D(-\Delta)^{z/2} N(x,t) = r  N(x,t).
\end{equation}
This critical diffusion equation essentially tells that, in the frame of the so-called one-group diffusion equation, moving from neutron transport theory (without feedback) to reactor physics theory (i.e. neutronics with feedback) mainly consists in replacing the standard Laplacian ($z=2$) by the fractional Laplacian with $z=1.901$. The equilibrium spatial distribution of the neutron field can therefore be obtained through the analysis of the first eigenmode of this fractional Laplacian. Since the reactor has a finite spatial extent, it is necessary to estimate this eigenmode in a bounded domain, keeping in mind that the critical exponents obtained through the use of the renormalization group are valid only for an infinite medium. To avoid the intricate analysis of critical exponents of the directed percolation in finite domain, we will anyway use this (infinite medium) dynamic critical exponent while keeping in mind that the results of the analysis will be only valid within the bulk of the medium (i.e. far from the boundaries). The precise topic of the properties of the fractional Laplacian in bounded domains has been investigated in the seminal paper of Zoia et al \cite{Zoia2007FractionalDomains}. This paper proposes different way to numerically tackle such equations. One possible -and elegant- way being to perform a Monte Carlo simulation of the neutron transport where the standard diffusion kernel is replaced by Lévy flights. Given the different hypotheses of our reactor model (diffusion instead of transport, no energy dependence, uniform medium, restriction to the bulk of the domain,...) we are only interested in approximate solutions, and therefore another possibility is to have recourse to approximate analytical expression as the one given by \cite{Kwasnicki2012EigenvaluesInterval} for example. Figure \ref{compa_z} reports the comparison between the exact solution of the critical diffusion equation that models neutron transport and such an approximate solution of our critical super-diffusion equation that models reactor physics. This approximate solution has the form
\begin{equation}
    \label{1DLévyFlight}
     N(x) = A \cos \Bigl( \left( \frac{\pi}{2}-(2-z)\frac{\pi}{8}\right) \frac{x}{L} \Bigr)
\end{equation}
with A being a normalization constant, and is obtained for a $1d$ uniform medium by imposing Dirichlet boundary conditions ($N(\pm L) = 0$) at the boundaries $L$ of the domain for $z=2$. For $z<2$, the neutron density at the boundaries does not vanish: this can be interpreted as long-distance flight bringing some neutrons to cross the boundary instead of being absorbed at it. But in the present case we are not interested in the properties of the medium close to the boundaries (where the dynamic critical coefficient is not valid anymore), hence the proper boundaries conditions for $z=1.901$ are not discussed in this paper. It is however worth noting that the topic of boundaries conditions for Lévy flights and their physics interpretation is currently under active investigation (see \cite{Constantin} for instance). 
Equation Eq.~\ref{1DLévyFlight} can also be extended to $3d$ medium by multiplication of the modes along each axis since  $N(x,y,z)=N(x)N(y)N(z)$ (this separation of variables stands for homogeneous partial differential equations in general \cite{Duderstadt1979TransportTheory}), making the study of a $1d$ medium meaningful. Also, for $z$ values close to $2$ the approximated solution comes close to being exact (the $L^2$ norm between the exact and the approximate solution is controled by $(2-z)$ \cite{Kwasnicki2012EigenvaluesInterval}). When $z=2$, this approximate solution becomes exact since one retrieves the usual solution of the critical diffusion equation $N(x) \sim \cos \Bigl( \frac{\pi}{2} \frac{x}{L} \Bigr)$. The $1d$ comparison between the spatial distributions of $z=2$ and $z=1.901$ (top plot of Figure \ref{compa_z}) finally reveals that, as could be qualitatively guessed, the super-diffusion slightly flatten the spatial distribution compared to diffusion. This macroscopic point of view emerges from the fact that the far-distance events characteristics of Lévy flights indeed tend to spatially redistribute the neutrons. In the frame of all the aforementioned hypotheses, the differences between the diffusion and the $z=1.901$ super-diffusion on local neutron densities are of the order of $1-2\%$ (bottom plot of Figure \ref{compa_z}), which is precisely the accuracy targeted by high precision reactor physics codes and methods, thus making these relative differences non-negligible.

\section{Conclusion}
\label{sec:conclusion}

The response functional formalism of Janssen and De Dominicis has proven to be a versatile and powerful companion for exploring the physics of branching processes. Applied to the well known equations of neutron transport in fissile media (neutronics), we have been able to recover with relative ease the main recent results of the field : moments of the neutron population within the reactor, effect of external neutron sources on the chain reaction, taking into account of delayed neutrons, and characterization of spatial and temporal correlations that may develop at low power. Along this path, the close connection between the statistical field theory of Janssen and De Dominics and  the coherent path integral method of Doi and Peliti has been underlined. This last approach is based on establishing a second quantized version of the neutron's branching process, and therefore relies on the solid-ground microscopic description of branching spatial Brownian motion that has been build in the past few decades. But the generality of the response functional approach lend itself to the inclusion of very general ad-hoc models whose microscopic origin would be difficult to implement. The restoring force that has been used in section \ref{sec:directed_percolation} is an example of such a heuristic approach. It allowed the study of a simplistic model of nuclear reactor, that couples neutronics to a prototype feedback model (the Doppler effect), which is usually not possible via the traditional master equations approach. Showing that the phase transition of the neutron population at criticality falls into the directed percolation universality class permitted to retrieve the critical exponents of the phase transition obtained thanks to the well established results of the renormalization group technics. \newline 

The main consequence of this work in the realm of nuclear reactor physics were evaluated by focussing on the corrected mean field equation of our simplified reactor physics model, using in particular the dynamic critical exponent $z$ previously obtained. In the same manner that this coefficient transforms the diffusion law $z=2$ into a super-diffusive Lévy flights with $z=1.901$ for $3d$ medium, the usual critical diffusion equation used to grasp the main features of neutronics was shown to be replaced, in this prototype model of reactor physics, by a critical super-diffusion equation. In the low feedback regime, and far from the boundaries, the main eigenmode of this critical super-diffusion equation was approximated using recent developments on the spectrum of the fractional Laplacian in bounded domains. In this frame, the solution of the standard mean field equation for the neutron local distributions might show deviations up to $1-2\%$, which invalidates many formal approaches and reactor physics analytical models (see \cite{Dellacherie, Tomatis} among many others) sometimes used to benchmark reactor physics codes, while questioning the accuracy targeted by the current high precision reactor physics codes themselves, which aim at calculating neutron fluxes with an accuracy below this precision (see \cite{Demaziere} for instance). \newline

Finally, to further assess the accuracy of mean field approaches (and hence reactor physics codes), it would be necessary to investigate the properties of the directed percolation in finite medium, while studying the effects of neutron energy (in the frame of a 2-group critical super-diffusion equation for instance) and improving the modeling of the feedback mechanism. A prominent role might also be played by the precursor of delayed neutrons. Indeed, these precursors being not subject to any feedback mechanism, they might inflect the system's dynamic into a different universality class.
Ultimately, reactor physics experiments aiming at measuring and detecting distortions of the neutron field could definitely sign this phenomenon of neutron percolation in reactor cores.

\section*{Acknowledgments}

The authors are thankful to L\'eonie Canet and Andrea Zoia for insightful discussions and comments. \\

\appendix

\section{Analytical calculation of the generalized propagators $\Sigma_{ab}$ ($a,b$ = $N,M$) in the two species model}
\label{app_a}

Focusing on the evaluation of $\Sigma_{NN}(t,t')$, and using the shorthand notation $a = r + \lambda_D$, the diagram of Eq.~\ref{sig_nn} translates into the infinite series, 
\begin{widetext}
\begin{equation}
    \Sigma_{NN}(t,t') = e^{ r (t - t')} \left\{  1 + \beta \lambda_D \int_{t'}^{t} dt_1 e^{-a t_1} \int_{t'}^{t_1} dt_2 e^{a t_2} +
    (\beta \lambda_D)^2 \int_{t'}^{t} dt_1 e^{-a t_1} \int_{t'}^{t_1} dt_2 e^{a t_2} \int_{t'}^{t_2} dt_3 e^{-a t_3} \int_{t'}^{t_3} dt_4 e^{a t_4} + ... \right\},
\end{equation}
or,
\end{widetext}
\begin{equation}
   \Sigma_{NN}(t,t') = e^{ r (t - t')} \left( 1 + \sum_{n \geq 1} (\beta \lambda_D)^{n}  A_{n}(t,t') \right),
\end{equation}
with, 
\begin{eqnarray}
   & & A_n(t, t') =\nonumber\\
   & & \hspace{-0.2 cm} \int_{t'}^t\int_{t'}^{t_1} ... \int_{t'}^{t_{2n-1}} \hspace{-0.1 cm} dt_1 ...  dt_{2n} e^{-a(t_1 -t_2 + ... +t_{2n-1} - t_{2n})}
\end{eqnarray}
One can first easily show that $A_n(t,t') = A_n(t-t',0)$, so that only $A_n(t) := A_n(t,0)$ needs to be evaluated. Then, the form of the exponential calls for a change of variable of the form $t_1, ..., t_{2n} \rightarrow t_1 -t_2, ..., t_{2n-1} - t_{2n}, t_{2n} = u_{1}, ..., u_{2n}$. Using this new set of variables, $A_n(t)$ can be written, 
\begin{equation}
     A_n(t) = \idotsint\limits_{ \substack{0 < u_1, \hdots, u_{2n}\\  u_1 + .. + u_{2n} < t}} du_1 \hdots  du_{2n}  e^{-a(u_1 + \hdots + u_{2n-1})},
\end{equation}
where the exponential now only depends on the odd terms $u_{2i+1}$. The constraint  $\sum_{i=1}^{2n}u_i < t$ can be enforced through the introduction of an indicator function $\mathbbm{1}_{\sum_{i=1}^{2n}u_i < t}$, so that, 
\begin{eqnarray}
\label{appA:int_to_split}
    & & A_n(t) =\nonumber\\
    & & \hspace{-0.2 cm}\idotsint\limits_{0 < u_1, \hdots, u_{2n}} du_1 \hdots  du_{2n}  e^{-a(u_1 + \hdots + u_{2n-1})} \mathbbm{1}_{\sum_{i=1}^{2n}u_i < t} 
\end{eqnarray}
The integration over even indices corresponds now to a volume integral over an $n$-dimensional simplex,
\begin{equation}
\label{vol_simplex}
     \idotsint\limits_{0 < x_1, \hdots, x_p} dx_1 \hdots  dx_p  \mathbbm{1}_{\sum_{i=1}^{p}x_i < \alpha} = \frac{\alpha^p}{p!}.
\end{equation}
Splitting Eq.~\ref{appA:int_to_split} over even/odd indices, one can then show that, 
\begin{widetext}
\begin{equation}
\begin{split}
A_n(t) & = \idotsint\limits_{0 < u_1, \hdots, u_{2n-1}} du_1 \hdots du_{2n-1}  \mathbbm{1}_{\sum_{i=0}^{n-1}u_{2i+1} < t}  e^{-a(u_1 + \hdots + u_{2n-1})}
\idotsint\limits_{0 < u_2, \hdots, u_{2n}} du_2 \hdots du_{2n}  \mathbbm{1}_{\sum_{i=1}^{n}u_{2i} < t - \sum_{i=0}^{n-1}u_{2i+1}} \\
& = \idotsint\limits_{0 < u_1, \hdots, u_{2n-1}} du_1 \hdots du_{2n-1}  \mathbbm{1}_{\sum_{i=0}^{n-1}u_{2i+1} < t}  e^{-a(u_1 + \hdots + u_{2n-1})} \frac{\left(t- u_1 - \hdots u_{2n-1}\right)^n}{n!}
\end{split}
\end{equation}
\end{widetext}
The integrand in the last expression only depends on the sum $u_1 + \hdots + u_{2n-1}$. Thus, defining $x = \sum\limits_{i=0}^{n-1}u_{2i+1}$ and changing the integration variables accordingly, one ends up with,
\begin{equation}
A_n(t) =  \int\limits_{0}^t  dx  e^{-ax} \frac{\left(t-x\right)^n}{n!} \frac{x^{n-1}}{(n-1)!}.
\end{equation}
$\Sigma_{NN}$ can thus be written,
\begin{eqnarray}
\Sigma_{NN}(t,0) & = &  e^{r t}
\Bigg\{ 1 +  \nonumber\\
  & & \hspace{-2 cm}  \left. \sum\limits_{n\geq0} (\beta \lambda_D)^{n+1} \int_0^t dx ~ e^{-ax}  \frac{\left(t-x\right)^{n+1}  x^{n} }{(n+1)! ~ n!} \right\}.
\end{eqnarray}
The latter expression involves the series development of the Modified Bessel function of the first kind $I_1$. Reinstating the lower bound $t'$ into the equation, one then has, 
\begin{eqnarray}
\Sigma_{NN}(t, t') & & =  e^{r (t-t')} \Bigg\{ 1 + \nonumber\\
& &  \int_0^{t-t'} d\tau  e^{-r\tau} \sqrt{\frac{\beta \lambda_D(t-t'-\tau)}{\tau}} \nonumber \\ 
\times & & I_1\left(2\sqrt{\beta \lambda_D \tau (t-t'-\tau)}\right)
\Bigg\}.
\end{eqnarray}
One the same model, one can show that,
\begin{eqnarray}
 \Sigma_{MM}(t,t') & & = e^{-\lambda_D (t-t')} \Bigg\{1 + \nonumber\\ 
 & &\int_{0}^{t-t'} d\tau  e^{a \tau} \sqrt{\frac{\beta \lambda_D(t-t'-\tau)}{\tau}} \nonumber\\
 & & \times I_1\left(2\sqrt{\beta \lambda_D \tau (t-t'-\tau)}\right) \Bigg\},
\end{eqnarray}
\begin{eqnarray}
     \Sigma_{NM}(t,t') & & = \beta  e^{-\lambda_d(t-t')} \nonumber\\
     & & \hspace{-1 cm}\times \int_{0}^{t-t'}  d\tau ~ e^{a\tau} I_0\left(2\sqrt{\beta \lambda_D \tau (t-t'-\tau)}\right),
\end{eqnarray}
and 
\begin{eqnarray}
 \Sigma_{MN}(t,t') & & = \lambda_D ~e^{r(t-t')} \nonumber\\
 & & \hspace{-1 cm}\times \int_{0}^{t-t'} d\tau  e^{-a\tau} I_0\left(2\sqrt{\beta \lambda_D \tau (t-t'-\tau)}\right).
\end{eqnarray}
All these expressions contain integrals that can be further evaluated. After successive change of variables, and setting $\Omega = 4 \beta \lambda_D + \lambda_D^2 + 2 \lambda_D r + r^2$, one can show \cite{gradshteyn} that, 
\begin{eqnarray}
    \Sigma_{NN}(t,t') &=& \frac{e^{ \frac{1}{2} (r - \lambda_D)(t - t')  }}{\sqrt{\Omega}} \left\{  \sqrt{\Omega} \cosh\left(  \frac{1}{2} \sqrt{\Omega} (t - t') \right) \right. \nonumber\\
    & & \left.  + (r + \lambda_D) \sinh\left(  \frac{1}{2} \sqrt{\Omega} (t - t') \right)  \right\},
\end{eqnarray}

\begin{eqnarray}
    \Sigma_{MM}(t,t') &=& \frac{e^{ \frac{1}{2} (r - \lambda_D)(t - t')  }}{\sqrt{\Omega}} \left\{  \sqrt{\Omega} \cosh\left(  \frac{1}{2} \sqrt{\Omega} (t - t') \right) \right. \nonumber\\
    & & \left.  - (r + \lambda_D) \sinh\left(  \frac{1}{2} \sqrt{\Omega} (t - t') \right)  \right\},
\end{eqnarray}

\begin{equation}
        \Sigma_{NM}(t,t') = 2 \beta \frac{e^{ \frac{1}{2} (r - \lambda_D)(t - t')  }}{\sqrt{\Omega}} \sinh\left(  \frac{1}{2} \sqrt{\Omega} (t - t') \right),
\end{equation}

\begin{equation}
    \Sigma_{MN}(t,t') = 2 \lambda_D \frac{e^{ \frac{1}{2} (r - \lambda_D)(t - t')  }}{\sqrt{\Omega}} \sinh\left(  \frac{1}{2} \sqrt{\Omega} (t - t') \right) .
\end{equation}


\bibliography{main,mendeley}

\end{document}